# Van der Waals thin films of WTe$_2$ for natural hyperbolic plasmonic surfaces

Chong Wang[1,2], Shenyang Huang[1,2], Qiaoxia Xing[1,2], Yuangang Xie[1,2], Chaoyu Song[1,2], Fanjie Wang[1,2], Hugen Yan[1,2]*

[1]State Key Laboratory of Surface Physics and Department of Physics, Fudan University, Shanghai 200433, China.

[2]Key Laboratory of Micro and Nano-Photonic Structures (Ministry of Education), Fudan University, Shanghai 200433, China

*E-mail: hgyan@fudan.edu.cn

## Abstract

A hyperbolic plasmonic surface supports highly directional propagating polaritons with extremely large density of states. Such plasmon polaritons have been realized in artificially structured metasurfaces. However, the upper bound of the achievable plasmon wave vector is limited by the structure size, which calls for a natural hyperbolic surface without any structuring. Here, we experimentally demonstrate a natural hyperbolic plasmonic surface based on thin films of $WTe_2$ in the light wavelength range of 16 to 23 microns by far infrared absorption spectroscopy. The topological transition from the elliptic to the hyperbolic regime is further manifested by mapping the iso-frequency contours of the plasmon. Moreover, the anisotropy character and plasmon frequency exhibit prominent temperature dependence. Our study demonstrates the first natural platform to host 2D hyperbolic plasmons, which opens exotic avenues for the manipulation of plasmon propagation, light-matter interaction and light emission in planar photonics.

**Introduction**

Hyperbolic plasmonic surfaces, whose iso-frequency contour in the wave vector space is a hyperbola, have been realized at visible[1], mid-infrared[2] and microwave frequencies[3] in metasurfaces, created by artificial sub-wavelength structuring or self-assembling carbon nanotubes. This enables a series of potential applications for planar photonics[4,5], including but not limited to nanoscale imaging[6,7], negative refraction[1] and enhancement of spontaneous emission[8]. Since the observation of highly confined and tunable plasmons in graphene[9-11], polaritons, especially plasmon polaritons[12] and phonon polaritons[13-15] in two-dimensional (2D) materials, have attracted substantial attention. Recently, 2D materials with in-plane anisotropic interplay of intraband and interband responses are predicted to naturally sustain hyperbolic plasmon polaritons (HPPs)[16,17]. In contrast to artificial surfaces, whose wave vectors in the hyperbolic regime are limited by the inverse of the structure size, natural hyperbolic surfaces support higher electromagnetic confinement and more diverging photonic density of states[18,19]. More importantly, the hyperbolic regime can be extended to mid-infrared and terahertz (THz) frequencies in 2D materials[19,20]. This range corresponds to energies of molecular vibrations and thermal radiation, promising opportunities in chemical sensing and thermal management.

While some anisotropic 2D materials, such as black phosphorous, are predicted to represent a natural class of HPPs[16,17], the experimental demonstration of any natural hyperbolic plasmonic surface has not been realized yet. In-plane hyperbolic phonon polaritons, on the other hand, have been shown in natural $MoO_3$ surfaces[13,21] and structured hexagonal boron nitride metasurfaces[22]. In comparison to phonon polaritons, plasmon polaritons exhibit inherently stronger coupling to light[23], making it more versatile for the management of thermal and spontaneous emission. Moreover, the hyperbolicity of plasmons in 2D materials can be tuned through chemical doping and gating[16], enabling active in-plane manipulation of polariton propagation.

Semimetal $WTe_2$ is one of the layered materials with in-plane anisotropic electrodynamics[24, 25]. Moreover, $WTe_2$ thin films host a wide range of remarkable electronic properties, such as extremely high mobility[26], tunable carrier density by electrostatic gating[27-29] and Mo doping[30], and being a candidate of type-II Weyl semimetals[31], which greatly facilitate the study of 2D plasmons. Previous reflection measurements on bulk $WTe_2$ crystals revealed temperature dependent anisotropic bulk plasma frequencies with extremely low optical scattering rates down to 0.25 $cm^{-1}$ at 6 K[25, 32]. Interestingly, the fitted in-plane dielectric functions exhibit a sign changing regime in the far-IR range[25], indicating the potential for realizing hyperbolic 2D plasmons in the thin films. However, the experimental observation of anisotropic plasmons modes in $WTe_2$ thin films and the defining evidence for the hyperbolic surface, the hyperbolic iso-frequency contour, are still absent. Especially, the hyperbolic regime is located at the far-IR range indicated by the bulk reflection data, where, on one hand, there is no readily available setup for near field experiments, and on the other, large area single crystal films are needed for far-field measurements, posing challenges in the observation of 2D plasmons.

Here, anisotropic 2D plasmons and hyperbolic plasmon dynamics are successfully observed in exfoliated $WTe_2$ thin films using Fourier transform infrared spectroscopy (FTIR), demonstrating anisotropic 2D material as an important platform to realize natural hyperbolic plasmonic surfaces.

**Results**

**Sample characterization and the scheme for transmission spectra measurements**

As shown in Fig. 1a, WTe$_2$ is crystallized in a T$_d$ type layered structure with a tungsten chain running along *a* axis. To maintain the lattice orientation, thin films were obtained through mechanical exfoliation from a single crystal (see Methods). Before investigation of the plasmon, we first explored the far-IR absorption spectra of an unpatterned thin film with thickness of about 60 nm on Si/SiO$_2$ substrate, as marked by dashed lines in Fig. 1c. The top edge is along *a* axis, which has been confirmed by Raman measurement (see Supplementary Fig. 1). Figure 1b shows the scheme for measuring absorption spectra, which is characterized by the extinction $1 - T/T_0$, where $T$ and $T_0$ are the transmission of the light through the film and the bare substrate, respectively. The amplitude of the extinction spectra is determined by the sheet optical conductivity $\sigma(\omega)$ [33] (see Methods), which is typically captured by the following expression[18]:

$$\sigma_{jj}(\omega) = \frac{i}{\pi} \frac{D_{jj}}{(\omega + i\Gamma)} + \frac{i}{\pi} \frac{\omega S_{jj}}{(\omega^2 - \omega_b^2 + i\omega\eta)} \qquad (1)$$

The conductivity is generally constituted of intraband (Drude response) and interband (bound states) electronic transitions in the form of the first and second terms on the right-hand side of equation (1), where index $j = a, b$, $D$ and $S$ denote the spectral weights, $\omega_b$ is the frequency of interband resonance, and $\Gamma$ and $\eta$ are the scattering widths for intraband and interband transitions, respectively. For simplicity, we retain only one Lorentz-type bound state in equation (1), which is sufficient for our discussion in the far-IR region. Electrons here are treated as normal carriers with Drude weight $D = \pi n_{\text{eff}} e^2 / m_{\text{eff}}$ [34], since the Fermi energy in undoped WTe$_2$ is well below the Weyl points[31]. Although two types of carriers[24, 35] are reported in WTe$_2$, to simplify the discussion, we treat them as one type with averaged effective carrier density ($n_{eff}$) and mass ($m_{eff}$).

**Anisotropic optical absorption in WTe$_2$ thin films**

Figure 1d displays the measured extinction spectra and corresponding fitting curves of the WTe$_2$ film in Fig. 1c. We can see that larger Drude response (dotted dashed lines) can be observed along *a* axis, while the interband component (dashed lines) is more intense along *b* axis. The different anisotropy of conductivity between the intraband and interband transitions, which tends to facilitate the formation of a relatively broad hyperbolic frequency range (see Supplementary note II for details), is further manifested by the polarization dependence of the fitted intraband and interband spectral weight (Fig. 1e and 1f). The mid-infrared absorption spectra exhibit consistent anisotropy, as shown in Fig. 1j (see also Supplementary Fig. 2 for spectra in a wider range). The fitted conductivity is summarized in the insert of Fig. 1d. The signs of the imaginary part of the optical conductivity along two principle axes determine the topology of the plasmonic surface[16, 36]. A frequency interval where $\sigma''_{aa} > 0$ (metallic electron response) and $\sigma''_{bb} < 0$ (dielectric response) can be found in the shaded area, giving a hyperbolic regime from 427 to 623 cm$^{-1}$. Using the fitted Drude weight, we can extract the ratio of effective mass by $m^b_{eff}/m^a_{eff} = D_{aa}/D_{bb} = 2.05$, consistent with the value of 2.2 measured in the bulk[25]. The temperature dependence of Fig. 1d was also investigated (see Supplementary Fig. 3). The Drude scattering width becomes broader at higher temperature (Fig. 1g), which has been attributed to Fermi liquid properties[32]. When the temperature increases, the

Drude weights along both axes are enhanced due to more thermal carriers, however with different enhancement ratios (Fig. 1h), leading to a less anisotropic effective mass at higher temperature (Fig. 1i).

**Anisotropic plasmon resonance modes in WTe₂ disk arrays**

When WTe₂ films are patterned into micro-structures such as micro-disk arrays (right inset of Fig. 2a), localized plasmons can be excited. Note that such patterning is only for plasmon detection in the far-field, rather than creating metasurfaces which need artificial structures much smaller than the plasmon wavelength to induce effective hyperbolicity. In the long wavelength limit, the 2D plasmon dispersion of free carriers is given by[34]

$$\omega_P = \left( \frac{e^2}{2\varepsilon_0 \varepsilon_{env}} n_{eff}/m_{eff} \right)^{\frac{1}{2}} \sqrt{q} \tag{2}$$

where $\varepsilon_0$ is the vacuum permittivity, $\varepsilon_{env}$ is the dielectric constant of the surrounding environment, $m_{eff}$ depends on the $q$ direction, and $q$ is the effective wave vector determined by the structure size. For ribbons with width $L$ and disks with diameter $d$, the effective $q$ is $\pi/L$ and $3\pi/4d$, respectively[37]. As shown in the lower insert of Fig. 2a, the WTe₂ film in Fig. 1c was etched into a disk array with diameter of $d = 5$ μm and spacing of $l - d = 3$ μm. When light was polarized along $a$ and $b$ axes, two distinctive plasmon modes were observed (Fig. 2a) at 10 K (see Supplementary Fig. 4 for plasmon modes with a 45° polarization). The splitting of those two modes is attributed to the anisotropic effective mass according to equation (2) (carrier density $n_{eff}$ is independent of direction). The effective mass ratio can be calculated by the resonance frequencies as $m_{eff}^b/m_{eff}^a = (\omega_P^a/\omega_P^b)^2 = 1.91$, in agreement with Drude response results for the unpatterned film. Similar to the Drude response, the plasmon modes are tunable by temperature as well, as shown in Fig. 2b. The fitted resonance frequency, width and the inferred mass ratio exhibit prominent temperature dependence (Fig. 2c – 2e). Two-peak structure can be observed at high temperature due to the hybridization with the polar phonon on $SiO_2$ substrates. More discussions on the temperature dependence and possible indication of Lifshitz transition can be found in Supplementary note IV.

**The hyperbolic regime derived from the anisotropic plasmon dispersion**

In order to investigate the plasmon dispersion at higher energy, where the plasmonic resonances are determined by both the intraband and interband transitions, we fabricated a set of rectangle arrays along the two optical axes of $WTe_2$ films on polycrystalline diamond substrates (see Methods). With such substrate, we avoided plasmon hybridization with substrate polar phonons. The films investigated on diamond substrates all have similar thickness of 100 ± 20 nm to maintain similar sheet optical conductivities. We first studied rectangle arrays with fixed aspect ratio $L_a = 2L_b$ to compensate the effect of anisotropic effective mass. The results of the extinction spectra along *a* and *b* axes are shown in Fig. 3a. A typical scanning electron microscopy (SEM) image of the rectangle array is shown in Fig. 3b. Two trends can be identified from Fig. 3a. The first trend is the plasmon dispersion. For large rectangles (8×4 $\mu m^2$), the plasmon modes along the two axes have nearly the same frequency (about 217 $cm^{-1}$). However, when the size of the rectangles is reduced, the plasmon frequencies along both axes blueshift but at different pace, leading to a splitting of the plasmon frequency (see Supplementary Fig. 7 for details). The second trend is the spectral weight. As plasmon moves to higher energies, the plasmon intensity along both axes gets reduced (with filling factors and film thickness taken into consideration), while plasmons along *b* axis have a larger intensity reduction rate (see Supplementary Fig. 7 for details).

The two trends can be further demonstrated by the plasmon spectra of a rectangle array with a higher aspect ratio in Fig. 3c. Although plasmons along the two optical axes have nearly identical frequency, the aspect ratio increases from 2 to 4, compared with the rectangle array (8×4 μm²) in Fig. 3a (the first pair of spectra from the top), further confirming the saturation tendency of plasmon frequency along *b* axis. In addition, we can see a clear decrease of intensity ratio between *b* and *a* axes in the two arrays (from 0.4 for 8×4 μm² to 0.13 for 1.8×0.45 μm²). The peak frequency of the latter array is closer to the hyperbolic regime (fitted below), which indicates a significant reduction of plasmonic spectral weight along *b* axis near the topological transition boundary between the elliptic and the hyperbolic regimes.

To study the plasmon dispersion in detail, all the measured plasmon frequencies of WTe$_2$ rectangle arrays on diamond substrates are summarized in Fig. 4a. Detailed plasmon absorption spectra of the red and blue spheres can be found in Supplementary Fig. 6. The dashed lines represent the standard 2D plasmon dispersion with $\omega \propto \sqrt{q}$. As shown in Fig. 4a, at low energy, the measured plasmon peaks follow the $\sqrt{q}$ scaling, thus it is legitimate for us to determine the effective mass anisotropy of free carriers from plasmons in Fig. 2. However, as the wave vector increases, the dispersion along both in-plane axes softens and departs from the purely free carrier case due to the coupling to interband transitions, and approaches an energy limit around 400 cm$^{-1}$ (600 cm$^{-1}$) for plasmons along *b* (*a*) axis at large wave vectors. Because of the lower limiting energy, the dispersion along *b* axis has a larger softening rate than that along *a* axis, leading to the frequency separation in Fig. 3a.

To fit the plasmon dispersion, the loss function $-\text{Im}(1/\varepsilon)$, defined from the imaginary part of the inverse of the dielectric function, is calculated by substituting equation (1) into the dielectric function with the following form in 2D case [34] (see Supplementary note I):

$$\varepsilon(q,\omega) = \varepsilon_{env} + \frac{i\sigma(\omega)}{\varepsilon_0 \omega} \cdot \frac{q}{2} \tag{3}$$

The fitting result is plotted as a pseudo-color map in Fig. 4a, which agrees well with the measured plasmon dispersion (see Methods), as indicated by the good match between the solid curves and data points. Figure 4b displays the imaginary parts of the conductivity extracted from the fitting of Fig. 4a. The hyperbolic regime is located at the shaded area in the range of 429 to 632 cm$^{-1}$ (15.8 to 23.3 microns in wavelength), fully consistent with the regime obtained through the optical conductivity of unpatterned film in Fig. 1d. The energies of $\sigma''_{jj}=0$, where $j = a$, $b$, are marked by the white dashed lines in Fig. 4a, which can be demonstrated to determine the energy boundaries of the plasmon dispersion along $a$ and $b$ axes (see Supplementary note II for details), implying that plasmon resonance modes can be found only along $a$ axis in the hyperbolic regime.

The hyperbolic regime can be further confirmed by plasmon intensity evolution in Fig. 4c, where plasmons along the two optical axes have intensity severely suppressed at corresponding boundaries of the hyperbolic regime. This is consistent with the prediction by the loss function, as shown by the intensity in the pseudo-color plot in Fig. 4a. A phenomenological coupled oscillator model, in which the plasmon and the bound state are represented by two coupled oscillators, is applied to give a quantitative description of the plasmon intensity evolution (see Supplementary note V for details). As shown in Fig. 4c, the agreement is good, manifesting that plasmons with nonzero intensity cannot be found along $b$ axis in the hyperbolic regime (the shaded area). This is consistent with the topology of hyperbolic plasmon iso-frequency contour, illustrated in the insert of Fig. 4a. The decreased plasmon weight is supposed to go to the interband plasmons at higher energies as predicted by the coupled oscillator model (see Supplementary note V for details). In fact, the coupling between intraband and interband transitions makes the bound states to have plasmon-like features. Such hybrid modes are termed interband plasmons[38]. Figure S10 shows the mid-infrared extinction spectra along $b$ axis of rectangle arrays with different $L_b$ length at 10 K. The frequencies of the two bound states shift to higher energies as the wave vector increases, providing an evidence for interband plasmons.

**Topological transition**

After establishing the energy range for hyperbolic plasmons, the iso-frequency contours of the plasmon at different energies are shown in Fig. 5b to exhibit the topological transition from the elliptic to the hyperbolic regime. The points in Fig. 5b with wave vectors $q$ away from the directions of principle axes were measured in ribbon arrays with a skew angle of $\theta$ with respect to $a$ axis (Fig. 5a), in which $q = \pi / L$ ($L$ is the ribbon width), with the direction perpendicular to the ribbon. The plasmon dispersion at each skew angle can be found in Supplementary Fig. 9. All the points in Fig. 5b can be fitted well by the solid curves based on the optical conductivity in Fig. 4b (see Methods). As shown in Fig. 5b, at energy below 429 cm$^{-1}$, which is in the elliptic regime, the iso-frequency contour is elliptic with the long axis running along the $b$ axis due to the smaller conductivity. At higher energy, the ellipse becomes flatter along $b$ axis, consistent with the increasing anisotropy of conductivities along the two principle axes, as shown in Fig. 4b. At energy above 429 cm$^{-1}$, where the imaginary parts of the conductivity have opposite signs along the two principle axes, the iso-frequency contours in

Fig. 5b become hyperbolic, which provide a defining evidence for the existence of a hyperbolic plasmonic surface.

**Discussion**

The fitted plasmonic resonance width at Fig. 2a is about 50 cm$^{-1}$ at 10 K along both axes and increased to about 400 cm$^{-1}$ at room temperature due to the Fermi liquid properties[32]. If we assume no inhomogeneous broadening, the lifetime is about 0.1 ps (0.013 ps) at 10 K (300 K). The low temperature lifetime is comparable to that of graphene plasmon on Si/SiO$_2$ substrate (about 0.05 to 0.1 ps)[39] but much lower than the values of plasmon polaritons in hBN encapsulated graphene (about 1 ps)[40] and phonon polaritons in hBN (about 2 ps)[41] and MoO$_3$ (about 8 ps)[13] films. At low temperature, the plasmon resonance width becomes broader in the hyperbolic regime with an average value of 100 cm$^{-1}$ due to the coupling with interband transitions. For WTe$_2$ films with thickness of 100 nm on diamond substrates, the calculated propagation length based on the plasmon lifetime has the maximal value of about 0.5 μm in the hyperbolic regime and decreases at smaller group velocities. In fact, such large resonance width in WTe$_2$ films is inconsistent with the low damping rate of the bulk plasma (down to 0.25 cm$^{-1}$)[32] and the high mobility (about 10,000 cm$^2$ V$^{-1}$ s$^{-1}$)[26] of bulk WTe$_2$. This is possibly due to the inhomogeneous broadening which causes an underestimation of the plasmon lifetime in our study. Moreover, the carrier scattering from the surface impurities and the charge inhomogeneity of the substrates[42] can cause plasmon broadening as well. However, these can be potentially alleviated by hBN encapsulation, as demonstrated in graphene[40].

It is noted that although the hyperbolicity is demonstrated in WTe$_2$ film with larger thickness (about 100 nm) than that of the widely studied monolayer 2D materials (e.g. grapheme, $MoS_2$), it can be treated as a real 2D plasmonic system since the plasmon wavelength is much larger than the film thickness. In fact, it is in principle possible to observe the same hyperbolic plasmons down to about 12 nm, above which the electronic structure is reported to remain the same as that of the thick film[43]. For few layer $WTe_2$ films, the band structures are dramatically modified due to finite-size effects, and it becomes a 2D topological insulator in monolayer limit[44]. It is interesting to check how the hyperbolic plasmons develop in few layer limit in future study.

Due to the relatively large film thickness, the light confinement factor (wavelength ratio between light and plasmon) is relatively low at low frequencies (about 6 at ~ 200 cm$^{-1}$ along $b$ axis in Fig. 3a), while this ratio is increased to about 40 at 512 cm$^{-1}$ (ribbon width 240 nm at a skew angle of 33 degree) within our measurements, which is comparable to the values in graphene plasmons (50-60 in Ref.[11] and 40 in Ref.[10]). It should be noted that such high value is realized in 100 nm thick film with the sheet optical conductivity of over one order of magnitude higher than that in graphene, manifesting the high electromagnetic confinement in the hyperbolic regime. The light confinement can be further increased by reducing the film thickness and reaches a maximum value of about 2300 at film thickness of ~10 nm. However, such factor will be reduced after considering the losses[36] and nonlocality[17], which might close the otherwise hyperbolic iso-frequency contour at large momentum. At high temperature, the light confinement capability will be weakened due to the increased carrier density.

The tunability is an unique characteristic for the hyperbolic plasmonic surface in 2D materials[16]. In fact, there have been many works reporting tunable electrical properties by electrostatic gating for WTe$_2$ films, some of which have a film thickness of above 10 nm[28, 29]. In this work, the plasmons are observed in WTe$_2$ film down to about 50 nm, and thinner films are needed in future work to enable the electrostatic gating. Moreover, the recently developed solid ion gating technique[45] provides a chance to tune the carrier density for relatively thick films. Besides, the chemical doping method is another possible way to tune the WTe$_2$ plasmons. For example, the carrier density and effective mass have been reported to be tuned in Mo$_x$W$_{1-x}$Te$_2$ single crystals by different Mo doping contents[30].

In conclusion, we use FTIR to demonstrate the existence of a natural hyperbolic plasmonic surface in exfoliated WTe$_2$ thin films in the far-IR range. The same kind of phenomena, in principle, could be observed in near field spectroscopy[13, 21, 22]. However, due to the restrictions of laser wavelength and cryogenic sample conditions, such experiment is challenging at this stage[46]. Our successful observation of 2D plasmons in films obtained by mechanical exfoliation will stimulate investigations of hyperbolic plasmons at low or even room temperature in other anisotropic 2D materials, such as black phosphorus, T$_d$-MoTe$_2$ and PtTe$_2$, and topological plasmons in other layered topological semimetals (e.g. ZrSiS, ZrTe$_5$).

# Methods

**Sample preparation and fabrication:** WTe$_2$ single crystal films were prepared by a standard mechanical exfoliation technique from bulk WTe$_2$ crystals (HQ Graphene) onto substrates. The typical film dimension is about 300 μm, greater than the beam size under an infrared microscope. The film thickness was determined by a Bruker Dimension Edge AFM system (probe model RTESP-300, tapping mode). Here, all the films we investigated have thickness of above 50 nm, in which the possible surface oxidation has negligible effect on the spectrum measurement[47]. Actually, no degradation of plasmonic devices was found during the nano-fabrication and measurement process. Two types of substrates were used in this work. One is Si/SiO$_2$ substrate with SiO$_2$ thickness of 300 nm. The other is polycrystalline diamond grown by chemical vapor deposition (CVD) method with thickness of 300 μm, in which no polar phonon exists in our range of interest. Using the latter substrate, one can eliminate the hybridization effect of surface polar phonons with plasmon and hence measure the intrinsic plasmon dispersion. Disk/rectangle/ribbon arrays were fabricated from the exfoliated films by electron-beam lithography and subsequent reactive ion etching. Sulphur hexafluoride (SF$_6$) was used as the reaction gas. For the rectangle (ribbon) arrays, the gaps between adjacent rectangles along $a$ and $b$ axes (ribbons) are kept larger than half of $L_a$ and $L_b$ (ribbon width), so that the rectangles (ribbons) can be approximately regarded as isolated resonance cavities. Because of the inevitable lateral etching, the dimensions listed in Fig. 3a are nominal values, with real length of about 10% uncertainty and the corresponding aspect ratio of 2 ± 0.15 (see Supplementary Table 1). The film thickness $d_{\text{film}}$ is normalized to 100 nm with the wave vector $q$ multiplied by $d_{\text{film}}/100$ in Fig. 4a and 5b, given that the sheet conductivity $\sigma$ is proportional to the thickness.

**Far-IR optical spectroscopy:** For the polarized far-IR measurements, we used a Bruker FTIR spectrometer (Vertex 70v) integrated with a Hyperion 2000 microscope and a liquid-helium-cooled silicon bolometer as detector. The incident light was focused on $WTe_2$ films with a 15x IR objective. A THz polarizer was used to control the light polarization. The low temperature measurements were performed in a helium-flow cryostat (Janis Research ST-300) with a pressure of about $1 \times 10^{-6}$ mbar. Throughout the entire measurements, nitrogen gas was purged to an enclosed space housing the cryostat. This procedure minimized the absorption of infrared light by moisture in air and effectively increased signal / noise ratio.

**Fitting of the extinction spectra of the bare film and disk/rectangle/ribbon arrays:** The extinction spectra are determined by the complex dynamic conductivity $\sigma(\omega)$:

$$1 - T/T_0 = 1 - \frac{1}{\left|1 + Z_0 \sigma(\omega)/(1+n_s)\right|^2} \tag{4}$$

where $Z_0$ is the vacuum impedance, $\omega$ is the frequency of light, and $n_s$ is the refractive index of the substrate. The dynamic conductivity contributed by the plasmon mode in a disk/rectangle/ribbon array is given by:

$$\sigma_P = i \frac{f \cdot S_P}{\pi} \frac{\omega}{(\omega^2 - \omega_P^2) + i\Gamma_P \omega} \tag{5}$$

where $\omega_P$ and $\Gamma_P$ are the frequency and resonance width of the plasmon, $S_P$ is the spectral weight, and $f$ is the filling factor ($WTe_2$ micro-structure area over the total area).

The extinction spectra of the bare film in Fig. 1d are fitted using equations (1) and (4).

Spectral weights $D$ and $S$, and the resonance widths $\Gamma$ and $\eta$ are determined along both axes. The frequency of the bound states $\omega_b$ is fixed to 800 cm$^{-1}$, which is between the frequencies of the two bound states in Fig. 1j.

Plasmonic extinction spectra in the disk, rectangle and ribbon arrays are fitted using equation (4) and (5), with the contribution of possible residual Drude response and bound states taken into account. For plasmons, spectral weight $S_P$, plasmonic resonance frequency $\omega_P$, and resonance width $\Gamma_P$ are fitting parameters. The fitted plasmonic resonance width of the disk array is 30% smaller than the Drude scattering width in the bare film. This is probably due to the fact that the measured spectra range at low energy side is limited by light intensity and available only above 100 cm$^{-1}$. Thus only a portion of the Drude spectrum is measured, compromising the fitting accuracy. This also explains why the singularity in Fig. 2c is not clear enough in the temperature dependence of Drude weight in Fig. 1h.

**Calculating the loss function and iso-frequency contours:** The pseudo-color plot in Fig. 4a shows the fitting result of loss function based on the dielectric function of equation (3) and the conductivity of equation (1) described in the main text. Spectral weights $D$ and $S$, and the resonance widths $\Gamma$ and $\eta$ are fitted along both principle axes. To simplify the calculation, only the interband transition with the lowest resonance energy of 710 cm$^{-1}$ is considered. The dielectric constant of environment $\varepsilon_{env}$ is set to 3.3. The solid black curves in Fig. 4a are plotted to represent the peak position of the maximum value of the loss function at each given wave vector. The fitted Drude weights are 8.08×10$^{11}$ and 4.49×10$^{11}$ Ω$^{-1}$·s$^{-1}$ along $a$ and $b$ axes, respectively. The mass ratio calculated from the Drude weight is about 1.8, consistent with the ones obtained from the bare film and disk array. The fitted interband transition weights are $4.31\times10^{11}$ and $8.07\times10^{11}$ $\Omega^{-1}\cdot s^{-1}$ along *a* and *b* axes, respectively. The weight ratio is about 1.87, with larger amplitude along *b* axis, which is consistent with the results in the unpatterned film. The fitted scattering width of the Drude modes and bound states are nearly identical along both axes, with value of about 60 cm$^{-1}$ and 170 cm$^{-1}$, respectively.

For a ribbon array with a skew angle of $\theta$ with respect to *a* axis, in which case the wave vector *q* is not along the principle axes, the loss functions are calculated by substituting the conductivity $\sigma = \sigma_{aa}\cos^2\theta + \sigma_{bb}\sin^2\theta$ into equation (3). See Supplementary Fig. 9 for the loss function calculation results and the measured plasmon dispersions in skew ribbon arrays. It should be noted that, in hyperbolic regime, the uniaxial surface supports the hybrid transvers-magnetic (TM) – transvers electric (TE) polaritons. However, in the nonretarded regime ($q \gg \omega/c$), which is the case for our work, the hybrid polaritons are dominated by the TM field components, which are the plasmon modes studied in this work[48].

At each skew angle, a line is plotted to represent the peak position of the maximum value of the loss function at each given wave vector, as shown in Supplementary Fig. 9. Thus, at a given frequency and skew angle, the wave vector length in the iso-frequency contour (lines of Fig. 5b) is determined by selecting the *q* value in the loss function fitting lines at corresponding skew angle and frequency.

## Data availability

The data that support the findings of this study are available from the corresponding author upon request.

## Acknowledgments

H.Y. is grateful to the financial support from the National Young 1000 Talents Plan, National Natural Science Foundation of China (Grant Nos. 11874009, 11734007), the National Key Research and Development Program of China (Grant Nos. 2016YFA0203900 and 2017YFA0303504), Strategic Priority Research Program of Chinese Academy of Sciences (XDB30000000), and the Oriental Scholar Program from Shanghai Municipal Education Commission. C.W. is grateful to the financial support from the National Natural Science Foundation of China (Grant No. 11704075) and China Postdoctoral Science Foundation. Part of the experimental work was carried out in Fudan Nanofabrication Lab.

## Author contributions

H.Y. and C.W. initiated the project and conceived the experiments. C.W. prepared the samples, performed the measurements and data analysis with assistance from S.H., Q.X., Y.X., C.S. and F.W.. C.W. carried out the fittings based on the loss function and coupled oscillator model. H.Y. and C.W. co-wrote the manuscript. H.Y. supervised the whole project. All authors commented on the manuscript.

## Competing financial interests

The authors declare no competing financial interests.

## Additional information

Correspondence and requests for materials should be addressed to H.Y. (hgyan@fudan.edu.cn).

# Figure

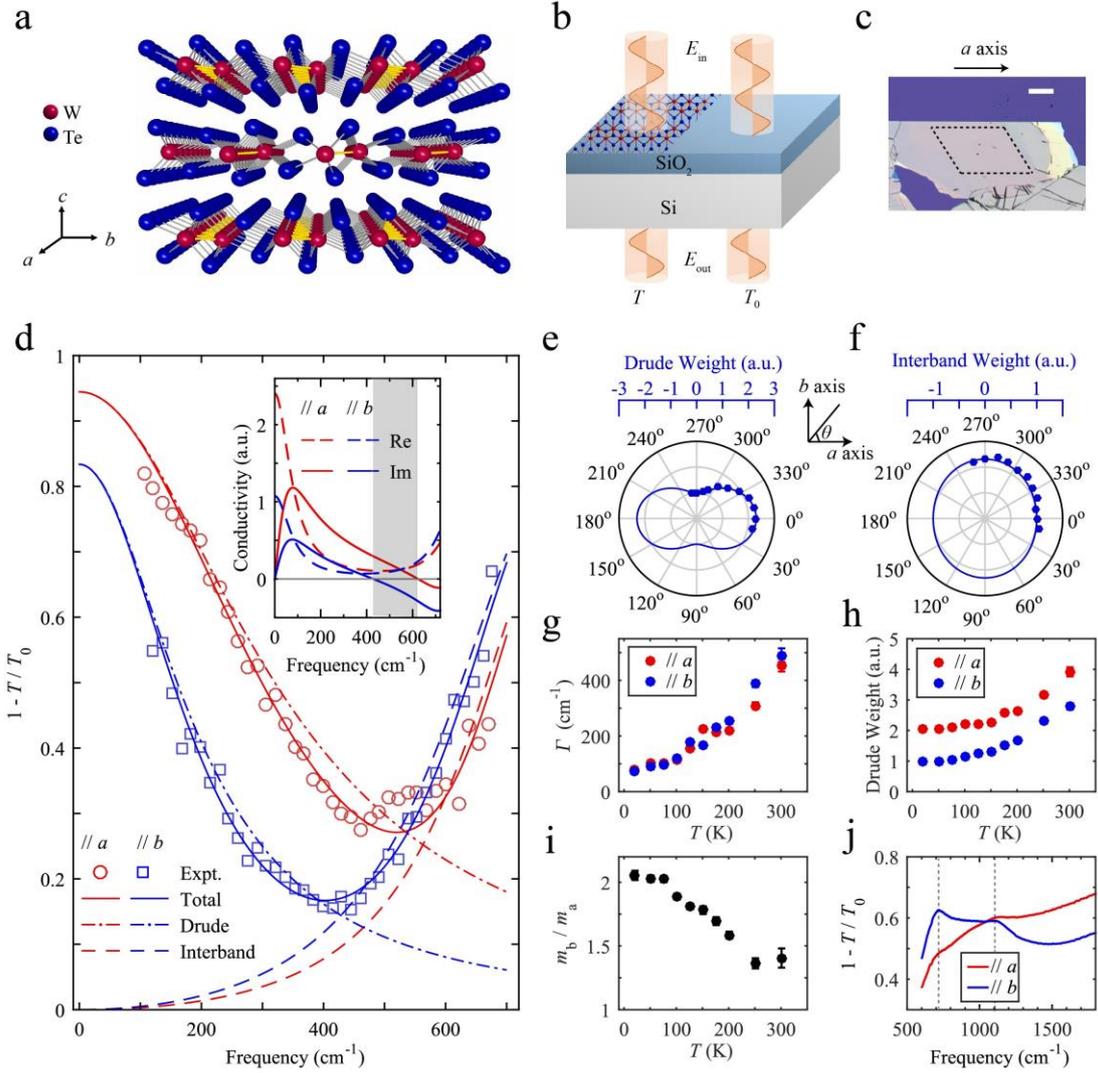

**Figure 1: Infrared absorption spectra in unpatterned exfoliated WTe$_2$ thin films. (a)** Schematic illustration of the layered crystal structure of WTe$_2$; the tungsten chains along *a* axis are shown as yellow zigzag segments. **(b)** Schematic of the setup for extinction spectra measurements. **(c)** Optical microscope image of the exfoliated WTe$_2$ thin film on Si/SiO$_2$ substrate. Scale bar is 100 μm. The optical measurements were performed in the uniform area marked by dashed lines, with thickness of about 60 nm. **(d)** Far-IR absorption spectra of the unpatterned exfoliated WTe$_2$ thin film in **(c)** with polarization along *a* and *b* axes at 20 K. Solid, dotted dashed and dashed lines are corresponding fitted curves of the total extinction spectra,

Drude component and interband transition component, respectively. Insert: Conductivity along *a* and *b* axes calculated by the fitting results. Shaded area represents the hyperbolic frequency regime. **(e)** and **(f)** Polarization dependence of the fitted Drude weight and interband weight. Curves are fitting results by $\cos^2\theta$. **(g)** and **(h)** Temperature dependence of the fitted scattering width and weight of Drude response along *a* and *b* axes. **(i)** Temperature dependence of effective mass ratio calculated by the fitted Drude weight. **(j)** Mid-infrared absorption spectra of an unpatterned $WTe_2$ thin film exfoliated on a polycrystalline diamond substrate with polarization along *a* and *b* axes. The film thickness is ~100 nm.

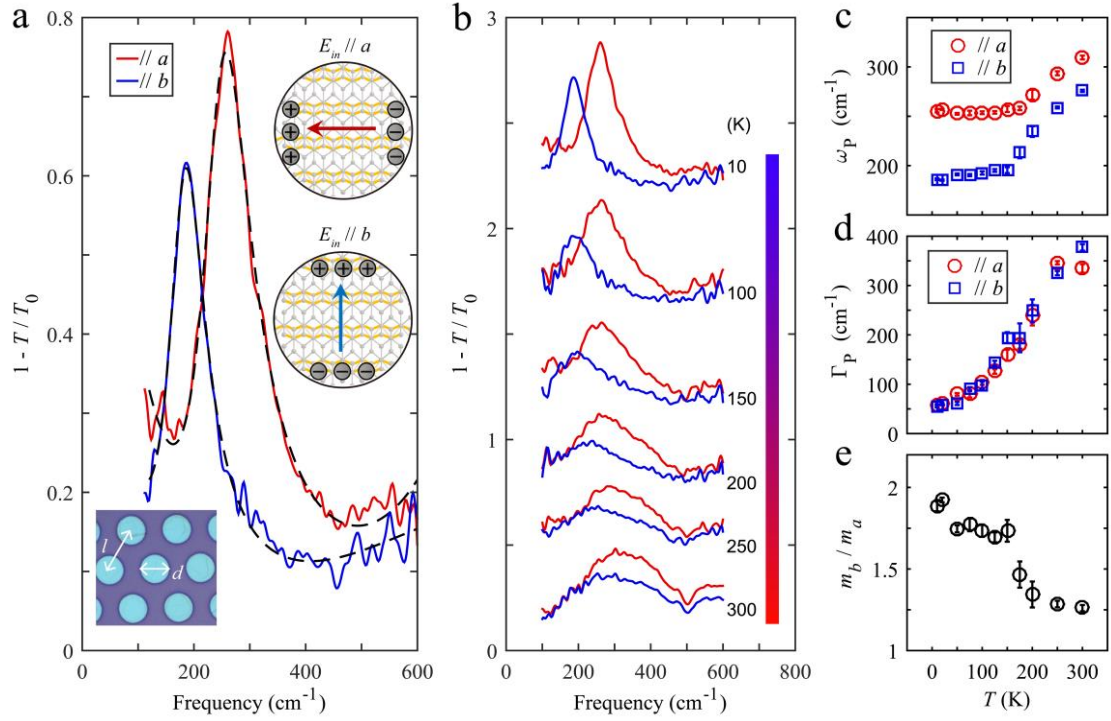

**Figure 2: Anisotropic plasmons in WTe$_2$ microdisk array with strong temperature dependence. (a)** Extinction spectra of the microdisk array fabricated on the film in Fig. 1c along $a$ and $b$ axes at 10 K, with Si/SiO$_2$ as substrates. The black dashed curves are the corresponding fittings (see Methods for the fitting procedure). The upturn of the spectra along $a$ axis at low energy is due to the Drude absorption from the unpatterned film outside the patterned area. Left insert: Optical microscope image of the microdisk array, with $l$ = 8 μm and $d$ = 5 μm. Right insert: Schematics of the plasmon resonance modes in the disk cavities with different polarizations. **(b)** Temperature dependence of the extinction spectra of the microdisk array with polarization along $a$ and $b$ axes. The spectra are vertically displaced for clarity. **(c)**, **(d)** and **(e)** Temperature dependence of the fitted plasmonic frequencies, resonance widths and the calculated effective mass ratio based on the plasmon spectra. All error bars are defined from fittings.

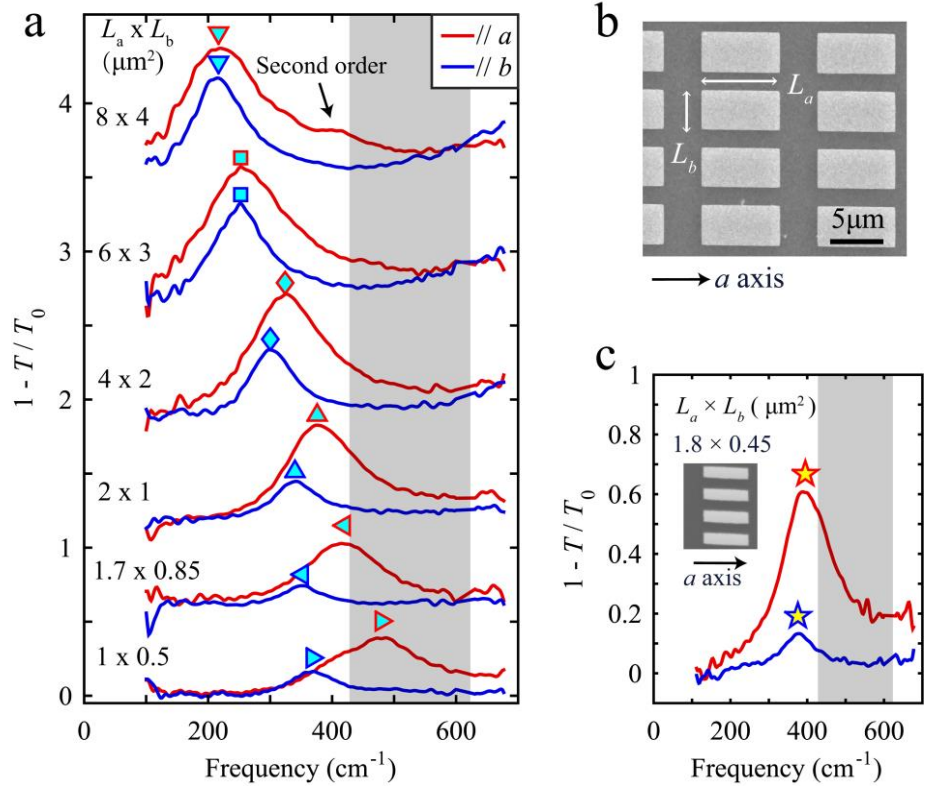

**Figure 3: Plasmon modes in the rectangle arrays on polycrystalline diamond substrates.**
(a) Extinction spectra of rectangle arrays with fixed aspect ratio $L_a/L_b = 2$ with polarization along $a$ and $b$ axes at 10 K. Offset for clarity. All the films have similar thickness of $100 \pm 20$ nm. (b) SEM image of a rectangle array with $L_a \times L_b = 8 \times 4$ μm². (c) Extinction spectra of the rectangle array ($1.8 \times 0.45$ μm²) with polarization along $a$ and $b$ axes. Insert shows the SEM image with scale bar of 1μm. The shaded area indicates the hyperbolic regime.

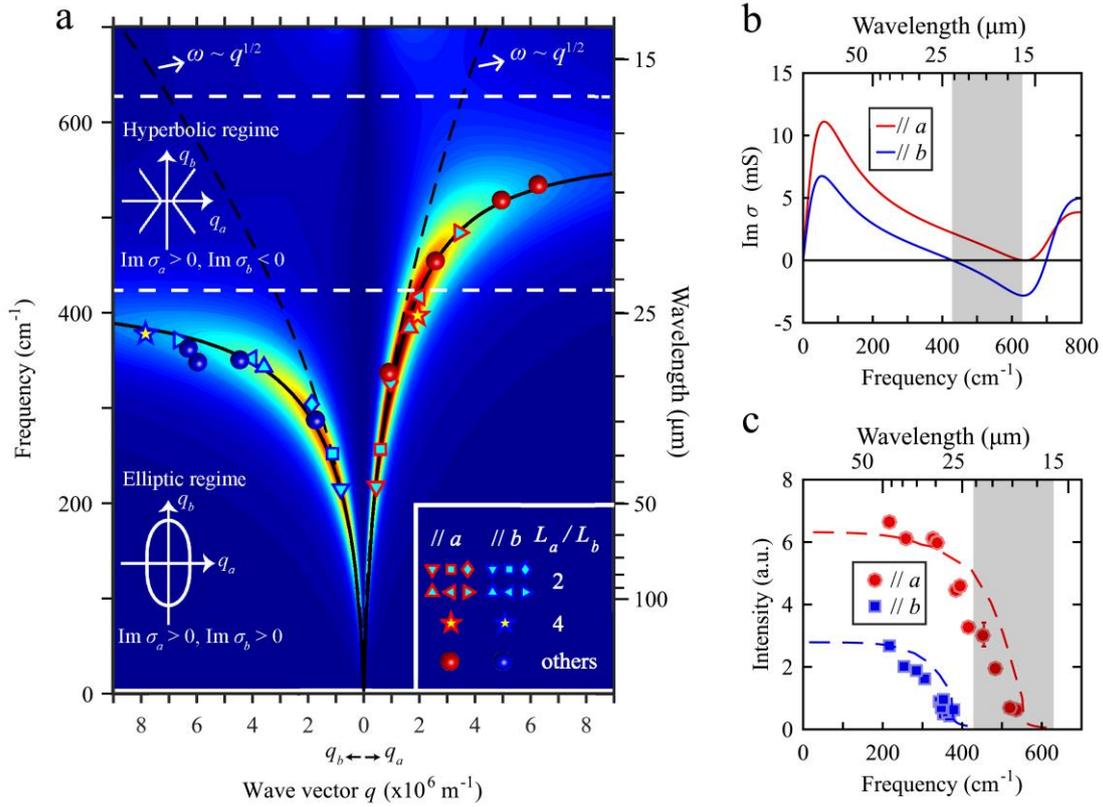

**Figure 4: Plasmon dispersion along the two principle axes.** (**a**) Plasmon dispersion for polarization along *a* and *b* axes measured in rectangle arrays. The calculated loss function is displayed as a pseudo-color map. The solid and dashed black curves represent the fitted dispersion with/without considering the interband transitions based on the loss function results. Dashed horizontal white lines are plotted to represent the hyperbolic regime obtained from the loss function calculation. Inserts are schematics of iso-frequency contours of the plasmon in the hyperbolic and elliptic regimes. See Supplementary note III for details. (**b**) Imaginary parts of the conductivity along *a* and *b* axes calculated by the loss function. (**c**) Intensity of plasmons along *a* and *b* axes in (**a**) as a function of the plasmonic resonance frequencies. Error bars are defined from fittings. The filling factors and film thickness have been taken into account. The dashed curves are fittings by the coupled oscillator model. All the shaded areas indicate the hyperbolic regime.

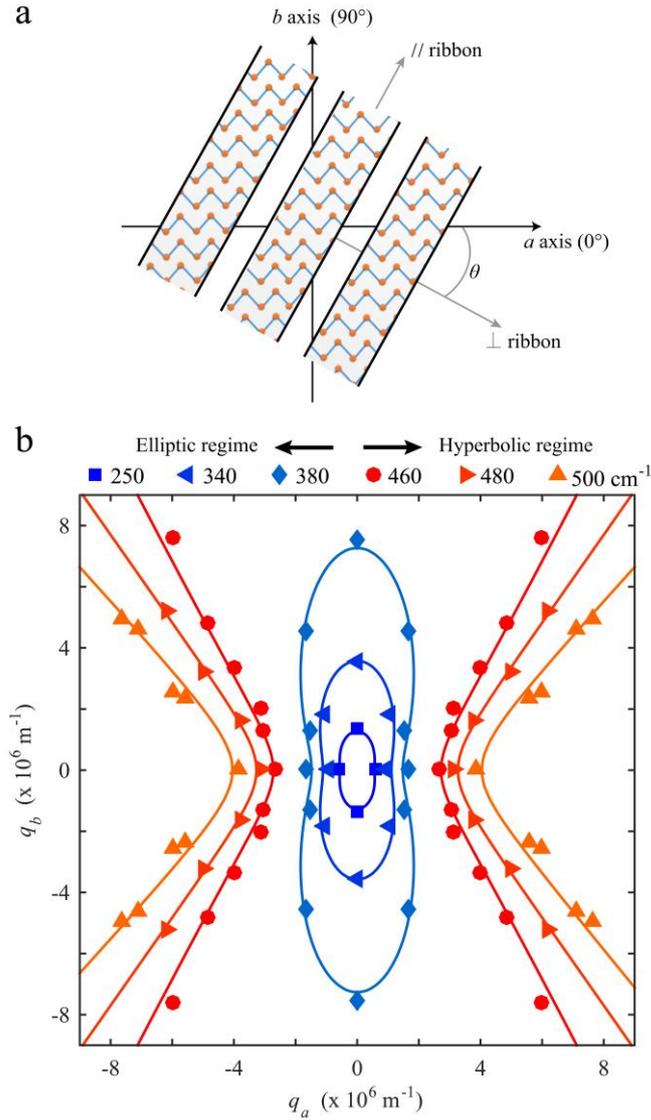

**Figure 5: Iso-frequency contours for the topological transition. (a)** The schematic of a skew ribbon array. Zigzag features are plotted inside the ribbons to represent the tungsten chains along $a$ axis. **(b)** Iso-frequency contours of the plasmon at different frequencies. Data points in the first quadrant are measured in rectangle/ribbon arrays of WTe$_2$ films with film thickness of $100 \pm 20$ nm on polycrystal diamond substrates with plasmon resonance frequencies at $250 \pm 10$ cm$^{-1}$, $340 \pm 10$ cm$^{-1}$, $380 \pm 10$ cm$^{-1}$, $460 \pm 8$ cm$^{-1}$, $480 \pm 8$ cm$^{-1}$ and $500 \pm 5$ cm$^{-1}$, respectively. Points in other quadrants are duplicated according to the crystal symmetry. The solid lines are plotted based on the optical conductivity (see Methods).